\documentclass[12pt,a4paper]{article}

\usepackage{graphicx}        

\usepackage{amsmath}
\usepackage{amssymb}

\title{Modified method of simplest equation and  exact traveling wave solutions of a hyperbolic reaction-diffusion equation}
\author{Ivan P. Jordanov$^{1,2}$, Nikolay K. Vitanov$^2$}
\date{$^1$ University of National and World Economy, Sofia, Bulgaria\\
	$^2$Institute of Mechanics, Bulgarian Academy of Sciences, Akad. G. Bonchev Str, 1113 Sofia, Bulgaria}

\begin{document}
\maketitle
\begin{abstract}
We discuss a class of hyperbolic reaction-diffusion equations and apply the modified method of simplest equation in order to obtain an exact solution of an equation of this class (namely the equation that contains polynomial nonlinearity of fourth order). We use the equation of Bernoulli as a simplest equation and obtain traveling wave solution of a kink kind for the studied nonlinear  reaction-diffusion equation.
	\end{abstract}
\section{Introduction}
Differential equations arise in mathematical analysis of many  problems from natural and social sciences \cite{bussenberg}, \cite{dim1} - \cite{dim7}, \cite{logan2}, \cite{e1}, \cite{e2}, \cite{vy1} - \cite{vy3}. The reason for this is that  the differential equations relate quantities with their changes and such relationships are frequently encountered in many disciplines such as  meteorology, fluid mechanics, solid-state physics, plasma physics, ocean and atmospheric sciences, mathematical biology, chemistry, material science, etc. \cite{debn,hirsch,murr,perko,strauss,verhulst,knvit,vsd2}. The qualitative features and mechanisms of many phenomena and processes in the above-mentioned research areas can be studied by means of exact solutions of the model nonlinear  differential equations. Examples are phenomena such as existence and change of different regimes of functioning of complex systems, spatial localization, transfer processes, etc. In addition the exact solutions can be used to test computer programs for numerical simulations. Because of all above exact solutions of nonlinear partial differential equations are studied very intensively \cite{ablowitz1,ablowitz2,ac,ames, benkirane,galakt, gardner,holmesx,kudr90,logan,leung, scott, strauss, tabor, vit09, vit09a}.The nonlinear PDEs are integrable or nonlitegrable. Well known methods exist for obtaining exact  solutions of integrable nonlinear PDEs , e.g., the method of inverse scattering transform or the method of Hirota  \cite{ablowitz2,ac,gardner,hirota,remos}.  Many approaches for obtaining exact special solutions of nonlintegrable nonlinear PDEs have been developed in the recent years, e.g., \cite{fan,he,kudr05x,herrem,vdk10,wazw1,wazw2}. In this chapter we shall use a version of the method of simplest equation called modified method of simplest equation \cite{kudr05x,kudr08,kudr9,vdk10,vit11,v15}. The method of simplest equation uses a procedure analogous to the first step of the test for the Painleve property \cite{kudr05, kudr07x,k10}. 
In  the   modified method of simplest equation \cite{vit10x,vd10,vit11a,v13,v13a,v14}  instead of this procedure (the procedure requires work in the space of complex numbers) one uses one or several balance equations.
The modified method of simplest equation has shown its effectiveness on the basis of  numerous applications, such as obtaining exact traveling wave solutions  of generalized Swift - Hohenberg equation and generalized Rayleigh equation \cite{vit11}, generalized Degasperis - Procesi equation and b-equation \cite{vit11a},extended Korteweg-de Vries equations \cite{v13,v15},generalized Fisher equation, generalized Huxley equation \cite{vit10x}, generalized Kuramoto - Sivashinsky equation, reaction - diffusion  equation, reaction - telegraph equation\cite{vdk10}, etc. \cite{v13a}, \cite{v14}, \cite{v17}.
\par
Below we shall discuss hyperbolic reaction-diffusion equation of the kind
\begin{equation}\label{eq1}
\tau \frac {\partial^2 Q}{\partial t^2}+\frac {\partial Q}{\partial t} = D    \frac {\partial^2 Q}{\partial x^2}+\sum_{i=1}^{n}\alpha_{i} Q^{i},
\end{equation}
where $Q=Q(x,t)$, $n$ is a natural number and $\tau$, $D$, and $\alpha_i, i=1, 2, ...,n$ are parameters. The difference between Eq.(\ref{eq1}) and the classic nonlinear reaction-diffusion equation is in the term $\tau \frac {\partial^2 Q}{\partial t^2}$ . Equation of class (\ref{eq1}) are known also as damped nonlinear Klein-Gordon equations \cite{alg,burq,gallay,gonzales,wang}. We note that reaction-diffusion equations have many applications for describing different kinds of processes in physics, chemistry, biology, etc \cite{cantrell,grz,wilh}. Traveling wave solutions of these equations are of special interest as they describe the motion of wave fronts or the motion of boundary between two different states existing in the studied system. Below we apply the modified method of simplest  equation (described in Sect.2) for obtaining
exact traveling solutions of nonlinear  reaction-diffusion PDE with polynomial nonlinearity of fourth order (Sect3).
The obtained waves are discussed in Sect. 3 and several concluding remarks are summarized in Sect.4.
\section{The modified method of simplest equation}
Below we shall apply the modified method of simplest equation.  The current version of the methodology used by
our research group is based on the possibility of use of more than one simplest equation \cite{v17}. We shall describe this version of the methodology and below we shall use the particular case when the solutions of the studied nonlinear PDE are obtained by use of a single simplest equation. The steps of the methodology are as follows.
\begin{enumerate}
\item By means of appropriate ans{\"a}tze (below we shall use a traveling-wave ansatz but in principle there can be one or several traveling-wave ans{\"a}tze such as $\xi = \alpha x + \beta t$; $\zeta =\gamma x + \delta t$, $\dots$. Other kinds of ans{\"a}tze may be used too) the solved  nonlinear partial differential equation is reduced to  a  differential equation $E$, containing derivatives of one or several functions
\begin{equation}\label{i1}
E \left[ a(\xi), a_{\xi},a_{\xi \xi},\dots, b(\zeta), b_\zeta, b_{\zeta \zeta}, \dots \right] = 0
\end{equation}
\item
In order to make transition to the solution of the simplest equation we assume that	any of the functions $a(\xi)$, $b(\zeta)$, etc.,  is a  function of another function, i.e.
	\begin{equation}\label{i1x}
	a(\xi) = G[f(\xi)]; \ \ b(\zeta) = F[g(\zeta)]; \dots
	\end{equation} 
\item
We note that the kind of the functions $F$ , $G$, $\dots$ is not prescribed. Often one uses a finite-series relationship, e.g., 
\begin{equation}\label{i2}
a(\xi) = \sum_{\mu_1=-\nu_1}^{\nu_2} q_{\mu_1} [f (\xi)]^{\mu_1}; \ \ \ 
b(\zeta) = \sum_{\mu_2=-\nu_3}^{\nu_4} r_{\mu_2} [g (\zeta)]^{\mu_2}, \dots 
\end{equation}
where $q_{\mu_1}$, $r_{\mu_2}$, $\dots$ are coefficients.
However other kinds of relationships may be used too. Below we shall work on the basis of relationships of kind (\ref{i2}). 
\item
The functions  $f(\xi)$, $g(\zeta)$ 
are solutions of simpler ordinary differential equations called simplest equations. For several years the methodology of the modified method of simplest equation was based on use of one simplest equation. The new version of the methodology allows the use of more than one simplest equation. 
The idea for use of more than one simplest equation can be traced back two decades ago to the articles of Martinov and Vitanov \cite{mv92a,mv92b,mv94}. 
\item
Eq.(\ref{i1x}) is substituted in Eq.(\ref{i1}) and let the result of this substitution be a polynomial containing $f(\xi)$, $g(\zeta)$, $\dots$. Next we have to deal with the coefficients of this polynomial. 
\item
A balance procedure is applied that has to ensure 
that all of the coefficients of the obtained polynomial of $f(\xi)$ and $g(\zeta)$ contain more than one term. 
This procedure leads to one or several balance equations for some of the parameters of the solved equation and for some of the parameters of the solution. Especially the coefficients $\nu_i$ from Eq.(\ref{i2}) as well as the parameters connected to the order of nonlinearity of the simplest equations are terms in the balance equations. Note that the coefficients of all powers of the polynomials have to be balanced (and not only the coefficient of the largest power). This is why the extended balance may require more than one balance equation.
\item
Eqs. (\ref{i1x}) represent a candidate for solution of Eq.(\ref{i1}) if all coefficients of the obtained polynomial of are equal to $0$. This condition leads to a system of nonlinear algebraic equations for the coefficients of the solved nonlinear PDE and for the coefficients of the solution. Any nontrivial solution of this algebraic system leads to a solution of the studied  nonlinear partial differential equation. Usually the system of algebraic equations contains many equations that have to be solved by means of a computer algebra system. 
\end{enumerate}
Below we shall search a solution of the studied equation of the kind
\begin{equation}\label{se1}
Q(\xi)=\sum_{i=0}^{n}a_{i}[\phi(\xi)]^i, \ \ \xi = x - vt
\end{equation}
where $\phi(\xi)$ is a solution of the Bernoulli differential equation
\begin{equation}
\label{se2}
\frac{d\phi}{d\xi}=a\phi(\xi)+b[\phi(\xi)]^k
\end{equation}
where $k$ is a positive integer. We shall use the 
following solutions of the Bernoulli equation
\begin{equation}\label{se3}
\phi(\xi)=\sqrt[k-1]{\frac{ae^{a(k-1)(\xi+\xi_0)}}{1-be^{a(k-1)(\xi+\xi_0)}}},\ \ \phi(\xi)=\sqrt[k-1]{-\frac{ae^{a(k-1)(\xi+\xi_0)}}{1+ be^{a(k-1)(\xi+\xi_0)}}}
\end{equation}
for the cases $b<0$, $a>0$ and $b>0$, $a<0$ respectively.
Above $\xi_0$ is a constant of integration.
\section{ Studied hyperbolic reaction-diffusion equation and application of the method}
Below we shall solve the equation
\begin{equation}\label{eq2}
\tau \frac {\partial^2 Q}{\partial t^2}+\frac {\partial Q}{\partial t} = D    \frac {\partial^2 Q}{\partial x^2}+\sum_{i=1}^{4}\alpha_{i} Q^{i}.
\end{equation}
Reaction-diffusion equation of this kind (with polynomial nonlinearity of fourth order) was used to model the propagation of wave fronts in populations systems
\cite{vit09,vit09a}. We apply the ansatz $Q(\xi ) = Q(x-vt)$
and then we substitute $Q(\xi$) by Eq.(\ref{se1}) where $n=2$, i.e.,
\begin{equation}\label{eq3}
 Q[\phi(\xi)] =  a_0 +  a_1 \phi(\xi) + a_2 \phi(\xi)^2.
\end{equation}
The balance procedure leads to a simplest equation of fourth order:
\begin{equation}\label{eq4}
\frac{d\phi(\xi)}{d\xi} = 
b_0 + b_1[\phi(\xi)]+ b_2[\phi(\xi)]^2 +b_3[\phi(\xi)]^3+ b_4[\phi(\xi)]^4.
\end{equation}
Above quantities $a_0$, $a_1$, $a_2$, $b_0$, $b_1$, $b_2$, $b_3$ and $b_4$ are parameters. We note here that we shall use a particular case of this simplest equation where $b_0=b_2=b_3=0$. The substitution of the traveling wave ansatz and Eqs.(\ref{eq3}),(\ref{eq4}) in Eq.(\ref{eq2})
leads to the following system of 9 algebraic equations:
\begin{eqnarray}\label{algsys}
&&10\,(D-v^2\tau)\,{a_2}\,{b_4}^{2} + \alpha_4\,{a_2}^{4}=0, \nonumber \\
&&18\,(D-v^2\tau)\,a_2\,{b_3}\,b_4=0,\nonumber \\
&&4\,\alpha_4\, a_0\,{a_2}^{3} + \alpha_3\,
{a_2}^{3} +(D-v^2\tau)\,(8\,{ a_2}\,{b_3}^{2} + 16\,{a_2
}\,{b_2}\,{b_4})=0, \nonumber \\
&&(D-v^2\tau)\,(14\,{a_2}\,{ b_1}\,{b_4} + 14\,
{a_2}\,{b_2}\,{b_3}) + 2\,v\,{a_2}\,{ b_4}=0, \nonumber \\
&&(D-v^2\tau)\,(12\, a_2\,b_1\, b_3+ 12\,a_2\, b_0\, b_4 + 6\, a_2\,{b_2}^{2})+ (3\,\alpha_3\,a_0\, +
\alpha_2 +\nonumber \\
&& 6\,\alpha_4\,{ a_0}^{2}\,)\,{a_2}^{2}
+ 2\,v\,{ a_2}\,{ b_3}=0, \nonumber \\ 
&&2\,v\,a_2\,b_2 +(D-v^2\tau)\,(10\,a_2\,b_0\,b_3 + 10\, a_2\,b_1\,{\it b_2})=0, \nonumber \\
&&3\,\alpha_3\,{a_0}^{2}\,{a_2} + 2\,\alpha_2\,{a_0}
	\,{a_2} + 4\,\alpha_4\,{a_0}^{3}\,{a_2} + (D-v^2\tau)\,(8\,
	{a_2}\,{b_0}\,{b2} + 4\,{a_2}\,{b_1}^{2}) + \nonumber \\
&& 2\,va_2b_1 + \alpha_1\,{a_2}=0, \nonumber \\
&&2\,v\,{a_2}\,{b_0} + 6\,(D-v^2\tau)\,{a_2}\,
{b_0}\,{ b_1}=0,\nonumber \\
&&2\,(D-v^2\tau)\,{ a_2}\,{b_0}^{2} + \alpha_4\,
{a_0}^{4} + \alpha_3\,{ a_0}^{3} + \alpha_1\,{a_0} + 
\alpha_2\,{a_0}^{2}=0.
\end{eqnarray}
A nontrivial solution of the  system (\ref{algsys}) is:
\begin{eqnarray}\label{sol}
b_1 &=& \frac{[ \alpha_3^3 (49 \alpha_3^3 \tau - 640 \alpha_4^2)]^{1/2}}{640 \alpha_4^2 D^{1/2}}
\nonumber \\
b_4 &=&  -\frac{7 \alpha_3^2 \tau a_2 (49 \alpha_3^3 \tau - 640 \alpha_4^2 ) \Bigg[\frac{\alpha_4^4 D a_2 \bigg(320 + \frac{49 \alpha_3^3 \tau - 320 \alpha_4^2}{\alpha_4^2} \bigg)}{\alpha_3^3 \tau (49 \alpha_3^3 \tau - 640 \alpha_4^2)} \Bigg]^{1/2}   }{80 \alpha_4^4 D \Bigg( 320 + \frac{49 \alpha_3^3 \tau - 320 \alpha_4^2}{\alpha_4^2}\Bigg)} \nonumber \\
b_0 &=& b_2 = b_3 = 0,
\nonumber \\
v &=& \frac{\alpha_4^2 D^{1/2} \Bigg( 320 + \frac{49 \alpha_3^3 \tau - 320 \alpha_4^2}{\alpha_4^2}\Bigg)}{7 \alpha_3 \tau [\alpha_3 (49 \alpha_3^3 \tau - 540 \alpha_4^2)]^{1/2}}, \nonumber \\
\alpha_2 &=& {\displaystyle \frac {3}{8}} \,{\displaystyle \frac {
		\alpha_3^{2}}{\alpha_4}} , \ \
\alpha_1 = {\displaystyle \frac {3}{64}} \,{\displaystyle 
	\frac {\alpha_3^{3}}{\alpha_4^{2}}}, \ \
a_0 =  -  \frac {1}{4}\, 
\frac {\alpha_3}{\alpha_4}, \ \  a_1 = 0 
\end{eqnarray}
Then the solution of the simplest equation becomes ($k=4$)
\begin{eqnarray}\label{se4}
\phi(\xi)&=&\Bigg\{\Bigg[\frac{[ \alpha_3^3 (49 \alpha_3^3 \tau - 640 \alpha_4^2)]^{1/2}}{640 \alpha_4^2 D^{1/2}} \exp \Big[3 \frac{[ \alpha_3^3 (49 \alpha_3^3 \tau - 640 \alpha_4^2)]^{1/2}}{640 \alpha_4^2 D^{1/2}} (\xi + \xi_0) \Big] \Bigg]  \nonumber \\
&&  {\Bigg /} \Bigg[ \Bigg[ 1 + \frac{7 \alpha_3^2 \tau a_2 (49 \alpha_3^3 \tau - 640 \alpha_4^2 ) \Bigg[\frac{\alpha_4^4 D a_2 \bigg(320 + \frac{49 \alpha_3^3 \tau - 320 \alpha_4^2}{\alpha_4^2} \bigg)}{\alpha_3^3 \tau (49 \alpha_3^3 \tau - 640 \alpha_4^2)} \Bigg]^{1/2}   }{80 \alpha_4^4 D \Bigg( 320 + \frac{49 \alpha_3^3 \tau - 320 \alpha_4^2}{\alpha_4^2}\Bigg)}  \Bigg]  \nonumber \\
&& \times  \exp \Big[3 \frac{[ \alpha_3^3 (49 \alpha_3^3 \tau - 640 \alpha_4^2)]^{1/2}}{640 \alpha_4^2 D^{1/2}} (\xi + \xi_0) \Big] \Bigg] \Bigg\}^{1/3},\nonumber \\
\phi(\xi)&=&\Bigg\{\Bigg[\frac{[ \alpha_3^3 (49 \alpha_3^3 \tau - 640 \alpha_4^2)]^{1/2}}{640 \alpha_4^2 D^{1/2}} \exp \Big[3 \frac{[ \alpha_3^3 (49 \alpha_3^3 \tau - 640 \alpha_4^2)]^{1/2}}{640 \alpha_4^2 D^{1/2}} (\xi + \xi_0) \Big] \Bigg]  \nonumber \\
&& {\Bigg /} \Bigg[ \Bigg[ 1 - \frac{7 \alpha_3^2 \tau a_2 (49 \alpha_3^3 \tau - 640 \alpha_4^2 ) \Bigg[\frac{\alpha_4^4 D a_2 \bigg(320 + \frac{49 \alpha_3^3 \tau - 320 \alpha_4^2}{\alpha_4^2} \bigg)}{\alpha_3^3 \tau (49 \alpha_3^3 \tau - 640 \alpha_4^2)} \Bigg]^{1/2}   }{80 \alpha_4^4 D \Bigg( 320 + \frac{49 \alpha_3^3 \tau - 320 \alpha_4^2}{\alpha_4^2}\Bigg)}  \Bigg]  \nonumber \\
&&\times \exp \Big[3 \frac{[ \alpha_3^3 (49 \alpha_3^3 \tau - 640 \alpha_4^2)]^{1/2}}{640 \alpha_4^2 D^{1/2}} (\xi + \xi_0) \Big] \Bigg] \Bigg\}^{1/3},\nonumber \\
\end{eqnarray}
for the cases $b_4<0$, $b_1>0$  and $b_4>0$, $b_1<0$  respectively. Thus the solutions of Eq.(\ref{eq2}) are
\begin{eqnarray*}
Q(\xi) &=& -  \frac {1}{4}\, 
\frac {\alpha_3}{\alpha_4} + a_2  \nonumber \\
&& \times \Bigg\{\Bigg[\frac{[ \alpha_3^3 (49 \alpha_3^3 \tau - 640 \alpha_4^2)]^{1/2}}{640 \alpha_4^2 D^{1/2}} \exp \Big[3 \frac{[ \alpha_3^3 (49 \alpha_3^3 \tau - 640 \alpha_4^2)]^{1/2}}{640 \alpha_4^2 D^{1/2}} (\xi + \xi_0) \Big] \Bigg] \nonumber \\
&&  {\Bigg /} \Bigg[ \Bigg[ 1 + \frac{7 \alpha_3^2 \tau a_2 (49 \alpha_3^3 \tau - 640 \alpha_4^2 ) \Bigg[\frac{\alpha_4^4 D a_2 \bigg(320 + \frac{49 \alpha_3^3 \tau - 320 \alpha_4^2}{\alpha_4^2} \bigg)}{\alpha_3^3 \tau (49 \alpha_3^3 \tau - 640 \alpha_4^2)} \Bigg]^{1/2}   }{80 \alpha_4^4 D \Bigg( 320 + \frac{49 \alpha_3^3 \tau - 320 \alpha_4^2}{\alpha_4^2}\Bigg)}  \Bigg]  \nonumber \\
&& \times \exp \Big[3 \frac{[ \alpha_3^3 (49 \alpha_3^3 \tau - 640 \alpha_4^2)]^{1/2}}{640 \alpha_4^2 D^{1/2}} (\xi + \xi_0) \Big] \Bigg] \Bigg\}^{2/3}, \nonumber \\
\end{eqnarray*}
\begin{eqnarray}\label{se5} 
Q(\xi) &=& -  \frac {1}{4} \frac {\alpha_3}{\alpha_4} + a_2  \nonumber \\
&& \times \Bigg\{\Bigg[\frac{[ \alpha_3^3 (49 \alpha_3^3 \tau - 640 \alpha_4^2)]^{1/2}}{640 \alpha_4^2 D^{1/2}} \exp \Big[3 \frac{[ \alpha_3^3 (49 \alpha_3^3 \tau - 640 \alpha_4^2)]^{1/2}}{640 \alpha_4^2 D^{1/2}} (\xi + \xi_0) \Big] \Bigg]  \nonumber \\
&& {\Bigg /} \Bigg[ \Bigg[ 1 - \frac{7 \alpha_3^2 \tau a_2 (49 \alpha_3^3 \tau - 640 \alpha_4^2 ) \Bigg[\frac{\alpha_4^4 D a_2 \bigg(320 + \frac{49 \alpha_3^3 \tau - 320 \alpha_4^2}{\alpha_4^2} \bigg)}{\alpha_3^3 \tau (49 \alpha_3^3 \tau - 640 \alpha_4^2)} \Bigg]^{1/2}   }{80 \alpha_4^4 D \Bigg( 320 + \frac{49 \alpha_3^3 \tau - 320 \alpha_4^2}{\alpha_4^2}\Bigg)}  \Bigg]  \nonumber \\
&& \times \exp \Big[3 \frac{[ \alpha_3^3 (49 \alpha_3^3 \tau - 640 \alpha_4^2)]^{1/2}}{640 \alpha_4^2 D^{1/2}} (\xi + \xi_0) \Big] \Bigg] \Bigg\}^{2/3}\nonumber \\
\end{eqnarray}				
for the cases $b_4<0$, $b_1>0$  and $b_4>0$, $b_1<0$  respectively.		
		\begin{figure}
			\includegraphics[angle=-90,width=0.7\linewidth]{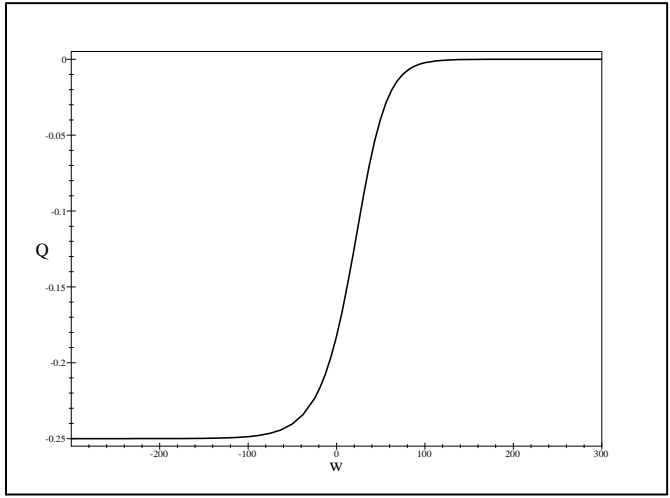}
			\caption
			{ Solution of equation (\ref{se5}). The values of
parameters are: $\tau=20$; $D=2$; $a_2=1 $; $\alpha_3=1$; $\alpha_4=1$; $\xi_0=0$. $w=\xi + \xi_0$.}
		\end{figure}
The obtained solutions (\ref{se5}) describe kink waves. Several of the waves are shown in Figs. 1-3. The parameters of the solutions are the same except the parameter $\alpha_3$
that has different values for the three kinks. As one can observe the decrease of the value of the parameter $\alpha_3$
leads to: (i) change of the values of $Q$ (from negative to positive); (ii) decrease of the width of the ink, and (iii) increase of the amplitude of the kink. Similar effects can be observed also in the case when the values of other parameters of the solution are varied.  		
	\begin{figure}
		\includegraphics[angle=-90,width=0.7\linewidth]{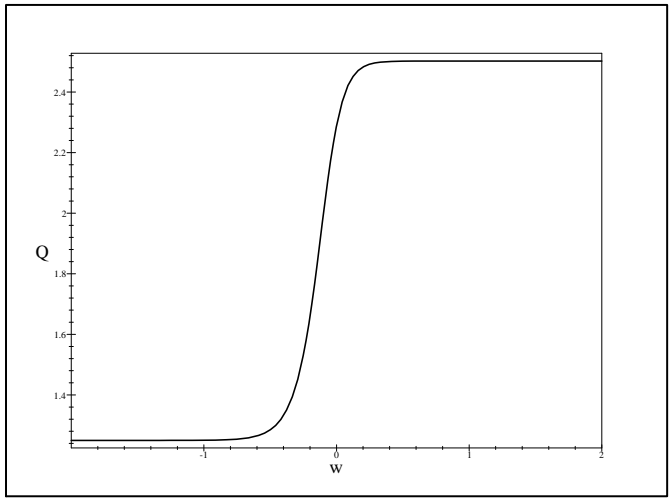}
		\caption{Solution of equation (\ref{se5}). The values of
			parameters are: $\tau=20$; $D=2$; $a_2=1 $; $\alpha_3=-5$; $\alpha_4=1$; $\xi_0=0$. $w=\xi + \xi_0$.}
	\end{figure}
	
	\begin{figure}
		\includegraphics[angle=-90,width=0.7\linewidth]{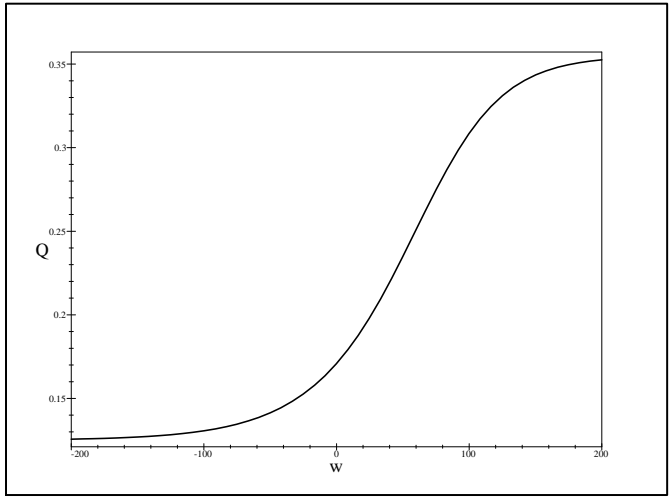}
		\caption{ Solution of equation (\ref{se5}). The values of
			parameters are: $\tau=20$; $D=2$; $a_2=1 $; $\alpha_3=-1/2$; $\alpha_4=1$; $\xi_0=0$.  $w=\xi + \xi_0$.}
	\end{figure}	
		\section{Concluding remarks}
In this chapter we have discussed the  nonlinear hyperbolic reaction-diffusion equation (\ref{eq2}). It  can be related
to the nonlinear reaction-diffusion equation that was used to model systems from population dynamics
\cite{vit09,vit09a}. We note that: 
\begin{itemize}
	\item The obtained solutions
of the hyperbolic reaction-diffusion equation do not contain as particular cases the kink solutions of the  rection-diffusion equation discussed in \cite{vit09,vit09a}. This is easily seen from the relationship for $b_4$ in Eqs.(\ref{sol}). If we set there $\tau =0$ then $b_4=0$ and we cannot construct a kink solution of the kind $Q(\xi) = b_1 + b_4 \phi(\xi)^2$. 
\item
Figs. 4 and 5 show the influence of increasing values of the parameter $\tau$ on the obtained kink solutions of the nonlinear hyperbolic reaction-diffusion equation.
\begin{figure}
	\includegraphics[angle=-90,width=0.7\linewidth]{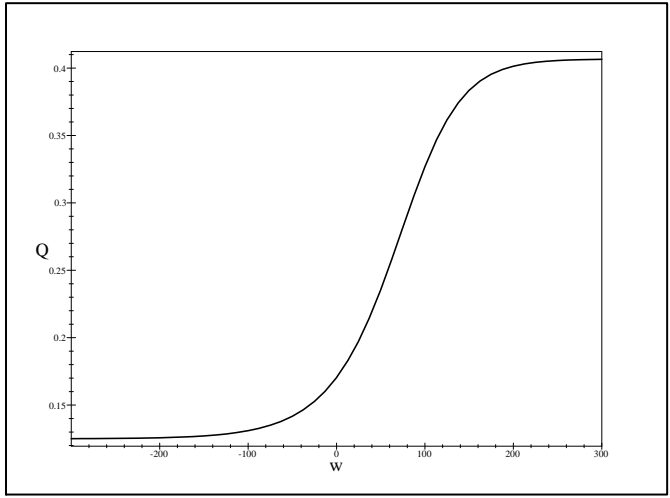}
	\caption{ Solution of equation (\ref{se5}). The values of
		parameters are: $\tau=10$; $D=2$; $a_2=1 $; $\alpha_3=-1/2$; $\alpha_4=1$; $\xi_0=0$.  $w=\xi + \xi_0$.}
\end{figure}	
\begin{figure}
	\includegraphics[angle=-90,width=0.7\linewidth]{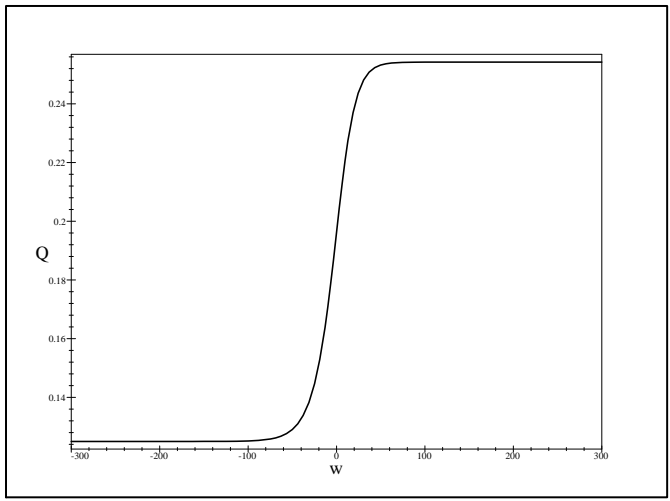}
	\caption
	{ Solution of equation (\ref{se5}). The values of
		parameters are: $\tau=1000$; $D=2$; $a_2=1 $; $\alpha_3=-1/2$; $\alpha_4=1$; $\xi_0=0$.  $w=\xi + \xi_0$.}
\end{figure}
As it can be seen from the figures the influence of the increasing value of $\tau$ on the kink profile is: (i) to decrease the amplitude of the kink, and (ii) to make the transition between the areas of lower and higher values of the kink more concentrated (i.e. this transition happens in the smaller interval of values of $w$).
\item
The exact solution of the studied nonlinear partial differential equation was obtained by means
of the modified method of simplest equation. We have shown that this method is an effective method for obtaining particular exact solutions of nonlinear partial differential equations  that do not belong to the class of integrable equations. 
\end{itemize}

\section*{Acknowledgements}
This study contains results, which are supported by the 
UNWE project for  scientific research with grant agreement No. NID NI -- 21/2016

\end{document}